\documentclass{aa}

\usepackage{graphicx}
\usepackage{epsfig}

\def\chem#1#2{$\rm{}^{#1}\kern-0.8pt#2$}
\def\reac#1#2#3#4#5#6{$\rm\,{}^{#2}\kern-0.8pt{#1}\,({#3}\,,{#4})\,
		      {}^{#6}\kern-0.8pt{#5}\,$}

\def\gsimeq{\,\,\raise0.14em\hbox{$>$}\kern-0.76em\lower0.28em\hbox  
{$\sim$}\,\,}
\def\lsimeq{\,\,\raise0.14em\hbox{$<$}\kern-0.76em\lower0.28em\hbox  
{$\sim$}\,\,}

\begin{document}
 
\thesaurus{06(02.14.1; 08.01.1; 08.09.2)}
\title{Uncertainties in the Th cosmochronometry}
\subtitle{}
 
\author{S.~Goriely\inst{1}\thanks{S.~Goriely is F.N.R.S. Research Associate.} 
\and B.~Clerbaux\inst{2}}

\offprints{S.~Goriely}
 
\institute{Institut d'Astronomie et d'Astrophysique, C.P. 226, Universit\'e
	   Libre de Bruxelles, B-1050 Brussels, Belgium\and Institut
	   Inter-Universitaire des Hautes Energies, C.P. 230, Universit\'e Libre
	   de Bruxelles, B-1050 Brussels, Belgium}

\date{Received April 20, 1999; accepted February 15, 1999}
 
\maketitle
%\markboth{ }{ }
 
\begin{abstract}
%______________________________________ Do not leave a blank line here!  
 Recent observations of r-nuclei, and in particular of Th, in ultra-metal poor
 stars revived the old idea that the Th cosmochronometry could provide an age
 estimate of the oldest stars in the Galaxy, and therefore a lower limit to the
 age of the Galaxy.  Unfortunately, some nuclear, astrophysics and observational
 uncertainties still affect the theoretical r-process models required to predict
 the original production of Th. The impact of these uncertainties on the
 prediction of the age of the Galaxy is analyzed and discussed.

\end{abstract}
 
\section{Introduction}
One of the methods called for estimating the age of the Galaxy is based on the
analysis of observed abundance of some long-lived radionuclides. The most
studied cosmochronometries involve $^{187}{\rm Re}$ or the actinides $^{232}{\rm
Th}$, $^{235}{\rm U}$ and $^{238}{\rm U}$. Though some promising results have
recently been achieved in decreasing the uncertainties affecting the $^{187}{\rm
Re}$ cosmochronometry, the predictions based on the trans-actinides can still be
regarded as relatively poor (Arnould \& Takahashi 1999). Nevertheless, the
recent observation of r-process elements, including Th, in ultra-metal-poor halo
stars, such as CS 22892-052 or HD 115444 (Sneden et al. 1996, 1998) has brought
some renewed excitement in the estimate of the age of the Galaxy on grounds of
the Th cosmochronometry (Cowan et al. 1997; Pfeiffer et al. 1998). With a
metallicity as low as [Fe/H] = --3 and a composition enriched in some pure
r-elements, these stars provide strong evidence that the production of heavy
elements by the r-process already took place early in the history of the
Galaxy. Moreover, the abundance pattern of the 15 r-elements heavier than Ba at
the surface of CS 22892-052 (or the 9 elements in HD 115444) shows a striking
similarity with the solar system r-abundance distribution, leading to the
tempting (though hazardous) conclusion that the r-process mechanism is "unique",
i.e any astrophysical event producing r-elements gives rise to a solar-like
abundance distribution. This conclusion has been critically analyzed by Goriely
\& Arnould (1997) who showed that this assumption may be valid indeed, but is by
far {\it not the only possible one}, as the observations in the limited $56\le Z
\le 76$ range are equally compatible with an abundance distribution that does
not fit the solar one outside the observed domain. This ambiguity is assigned to
the fact that the observed CS 22892-052 pattern of abundances reflects primarily
nuclear physics properties, and not one or another specificity of a blend of
r-process events. This universality assumption is a fundamental prerequisite to
build a Th cosmochronometry upon the abundance analysis of metal-poor stars at
the present time. In principle, it could be possible to derive the abundance of
Th ingested in these metal-poor stars from theoretical extrapolations based on
direct fits to the observed abundances. However, in practice, this exercise is
affected by uncertainties even greater than when basing the fits on the solar
abundances, because of the restricted number of elements observed, the
impossibility to distinguish isotopic ratios and the smaller precision in the
abundance determination compared with the data available in the solar system. In
particular, Goriely \& Arnould (1997) showed that the r-elements distribution at
the surface of CS 22892-052 could be reproduced satisfactorily by a random
superposition of canonical r-process events. In this case, the theoretical
extrapolation to the actinide region based on parametric r-process models is
simply meaningless. Nevertheless, future accurate observations of r-elements in
ultra-metal-poor star could change this situation.

For this reason, we will consider in the present paper the universality
assumption to be valid in order to analyze if, despite this difficulty, the
recent accurate observation of Th at the surface of ultra-metal-poor stars can
indeed provide a reliable estimate of the stellar age by comparing it with the
universal r-abundance of Th. Such a procedure requires the estimate of the Th by
r-process models, which are known to suffer from very many astrophysics and
nuclear physics problems, in spite of much recent theoretical and experimental
effort. In this respect, the Th problem is particularly acute, since with U, Th
is the only naturally-occuring nuclide beyond $^{209}{\rm Bi}$, so that the
estimate of Th production relies on extrapolation procedures based on fits to
the solar (or stellar) r-abundance distribution. In Sect. 2, a brief description
of the adopted r-process models is given in relation to the Th
cosmochronometry. In Sect. 3, the various uncertainties affecting the Th
production are studied and their impact on the estimate of the stellar age is
analyzed. In Sect. 4, it is shown by comparing the solar fits to the stellar
r-element distribution observed that future observations of Pb, Bi or U could
put the Th cosmochronometry on safer grounds.

\section{Th cosmochronometry and the r-process}

Assuming that the whole r-abundance distribution observed in CS 22892-052 and
HD115444 is essentially solar, it is straighforward to relate the star age $T_*$
to the Th abundances,
\begin{equation}
\Bigl({{\rm Th} \over {\rm Eu}} \Bigr)_{obs} = \Bigl({{\rm Th} \over {\rm Eu}}
\Bigr)_{r}
\exp\bigl[- T_* / \tau({\rm Th})\bigr]
\label{eq1}
\end{equation}

\noindent where $\tau({\rm Th})=20.27~{\rm Gyr}$ is the characteristic 
$\alpha$-decay
timescale of Th and the subscripts $obs$ and $r$ refer to the observed and
universal r-process abundance ratios, respectively. As classically done, the Th
abundance is here expressed relative to the spectroscopically relevant Eu
r-dominant element. The recent accurate observation of Th at the surface of CS
22892-052 amounts to $\log ({\rm Th}/{\rm Eu})_{obs}= -0.70\pm 0.08$ (Sneden et
al. 1996; Cowan et al. 1997). Th has also been observed at the surface of HD
115444, but its precise abundance remains to be determined (Pfeiffer et
al. 1998). Assuming that a solar-like mix of the r-elements ingested in these
halo stars originates from a small number of nucleosynthetic events that took
place just before the formation of the stars, the age of the star can be
estimated from Eq.~(\ref{eq1}) without calling for a complex model of the
chemical evolution of the Galaxy. The only difficulty of the methodology is
therefore related to the theoretical estimate of the r-production ratio $({\rm
Th}/{\rm Eu})_{r}$.

Unfortunately, the r-process remains the most complicated nucleosynthetic
process to model from the astrophysics as well as nuclear physics point of view
(for a review see Arnould \& Takahashi 1999). On the nuclear physics side, the
nuclear structure properties (such as the nuclear masses, deformation, \dots) 
of thousands of nuclei located between the  valley of $\beta$-stability and
the neutron drip line have to be known, as well as their interaction properties,
i.e the ($n,\gamma$) and ($\gamma, n$) rates, $\alpha$- and $\beta$-decay
half-lives and the fission probabilities.  Despite much recent experimental
effort, those quantities for most of the nuclei involved in the r-process remain
unknown, so that they have to be extracted on theoretical grounds and are
subject to the associated uncertainties.  On top of these nuclear difficulties,
the question of the astrophysical conditions under which the r-process can
develop is far from being settled. The site(s) of the r-process is (are) not
identified yet, all the proposed scenarios facing serious problems. For this
reason, only parametric approaches, such as the so-called canonical model
(Seeger et al. 1965) can be used to estimate the Th production. We use in the
present study the multi-event model\footnote{Note that, in the case of
r-processes responsible for elements observed in ultra-metal-poor stars, the
denomination "multi-event" does not refer to numerous astrophysical events, such
as supernova explosions, but rather to numerous components of a given
astrophysical event characterized by different thermodynamic conditions, for
example in the different layers of a given supernova.} (Bouquelle et al. 1996;
Goriely \& Arnould 1996) in which the best fit to the solar abundances is
derived from a superposition of canonical events with the aid of an iterative
inversion procedure. Compared with other treatments of the canonical model (e.g
Pfeiffer et al. 1998), a major advantage of the multi-event approach is to
provide an efficient tool for a systematic study of the various uncertainties
affecting the model (Goriely 1999). The iterative inversion method works in such
a way that the modification of a given (nuclear or astrophysics) input in the
r-process model leads to an automatic renormalization of the thermodynamic
conditions necessary to optimize the fit to the solar r-abundance distribution.
Therefore, the uncertainties affecting the input data of the parametric model,
as well as their impact on the Th production can be studied systematically
within the multi-event approach, as shown in the next section.

Our standard calculation is performed under the following thermodynamic
conditions: $1.3 \le T[10^9{\rm K}] \le 1.7$, $10^{22} \le N_n [{\rm cm}^{-3}]
\le10^{29}$ and $10
\le n_{cap} \le 200$ (where $T$ is the temperature, $N_n$ the neutron density
and $n_{cap}$ the number of neutrons captured per seed nucleus). Note that the
r-process calculations are performed making use of the waiting point
approximation, since under the thermodynamic conditions considered here, an
almost complete $(n,\gamma)-(\gamma,n)$ equilibrium is established (Goriely \&
Arnould 1996).  When not available experimentally, the nuclear data are taken
from the ETFSI nuclear masses of Aboussir et al. (1995) and from the gross
theory (GT2) of $\beta^-$ decay (and $\beta$-delayed neutron emission) of
Tachibana et al. (1990). In addition, $\alpha$-decay and fission processes are
also considered (before and after the neutron irradiation freeze-out). The
fission processes include spontaneous, $\beta$-delayed and neutron-induced
fission, the probabilities of which are calculated according to the
prescriptions of Kodoma \& Takahashi (1975) with the ETFSI fission barriers
(Mamdouh et al. 1998).  The procedure used to fit the solar r-abundance
distribution is similar to the one described in Bouquelle et al. (1996) though
each isotope is now given a weight inversely proportional to the error affecting
its solar r-abundance (Goriely, 1999). Since we are mainly concerned with Th
cosmochronometry, no details are given for the representative thermodynamic
conditions required to fit the solar system r-abundance distribution (such
details can be found in Goriely and Arnould, 1996).

\section{Uncertainties in the predicted Th abundance}

The r-process production of Th is obviously model dependent. However, for
cosmochronological purposes, it is of fundamental importance to know to what
extent the remaining uncertainties in the r-process modelling can affect the Th
synthesis.  From Eq.~(\ref{eq1}), it can be seen that the Th abundance has to be
determined within less than 16\% if we hope to predict the age of star within
less than 3 Gyr. A high accuracy has already been achieved observationally with
errors reduced to $\log \epsilon = 0.08$ affecting the stellar age by about 3.7
Gyr. Unfortunately, other uncertainties still need to be solved. These mainly
concern the r-process modelling, but before focussing on this subject, it is of
interest to stress that the normalization to the Eu abundance is not free from
uncertainties. As shown by Goriely (1999), even if the r-process models would be
able to reproduce exactly the Eu solar r-abundance, this value is still
uncertain by about 20\% (the observed abundance in the ultra-metal-poor stars is
known within 35\%) leading to an error in $T_*$ of about 6.6 Gyr. Note that with
respect to normalization procedure, it might be safer to use the Ho abundance,
since Ho is made of one stable isotope only and the s-contribution to its solar
abundance is even smaller than for Eu, i.e the error bars on its solar abundance
(about 13\%) are smaller than in the Eu case.  The additional uncertainties
related to the predicted Eu r-abundance are neglected in the present study by
normalizing the calculated Th abundance to the solar r-abundance of Eu.  The
complete absence of correlation in the production of Th and Eu in the canonical
approach of the r-process justifies this choice.

The sensitivity of the calculated r-process abundances, and in particular of the
Th abundance, to the different crucial inputs used in the multi-event model is
now examined and the impact of such uncertainties on the age of the CS 22892-052
star ($T^{CS}_*$) is discussed.

\subsection{Sensitivity to astrophysics conditions}

\begin{figure*}
% POSTSCRIPT FILE: F1.ps
  
\resizebox{\hsize}{!}{\includegraphics[scale=.7]{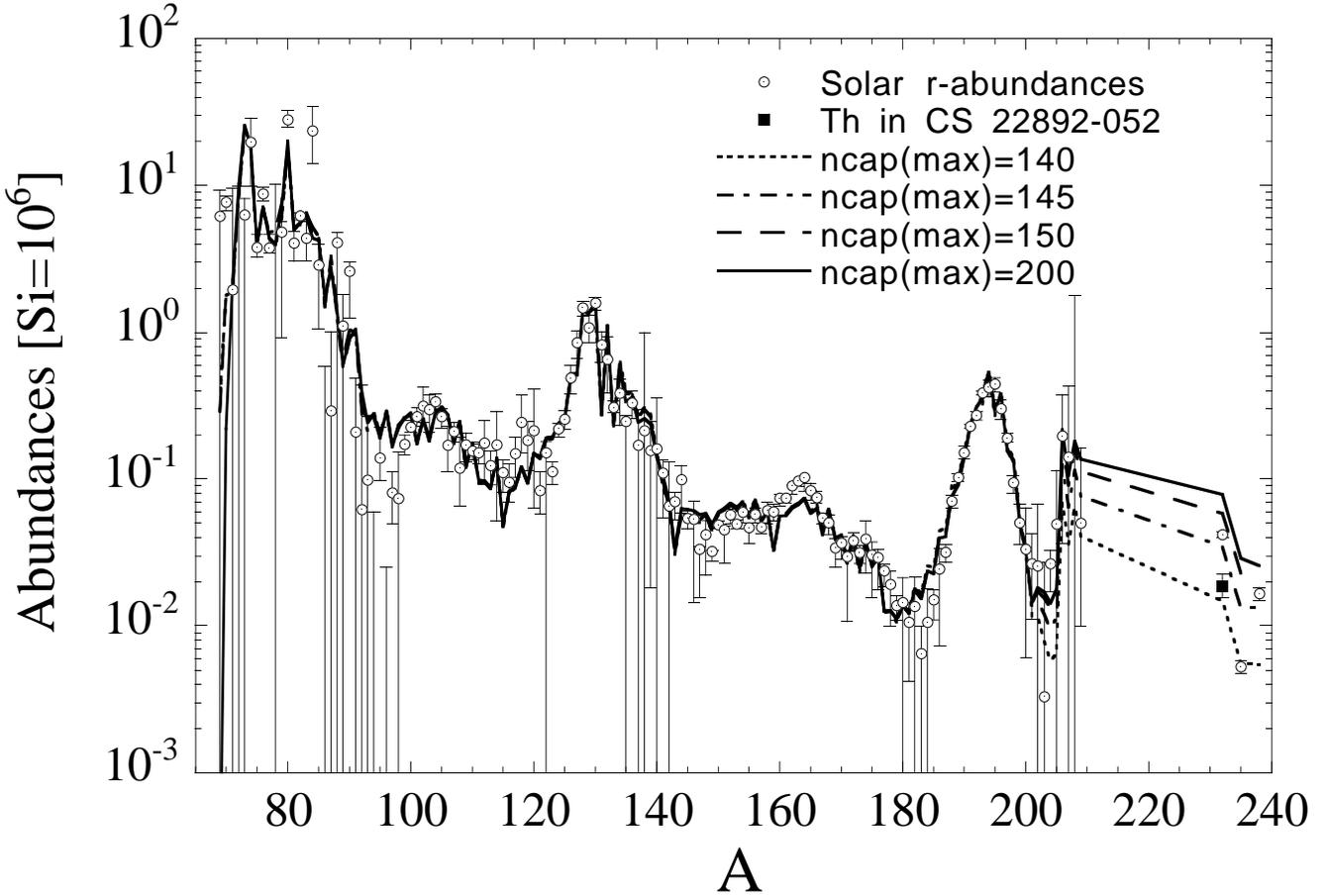}}
   \caption{ Comparison between the solar system r-abundances (Goriely 1999) and
   the distribution predicted by our standard multi-event superposition of
   events characterized by maximum values of $n_{cap}^{max}=$140 (dashed), 145
   (dot-dash), 150 (dotted) and 200 (full curve). The square corresponds to the
   Th abundance observed in CS 22892-052 (Sneden et al. 1996). The Bi and Th
   abundance are connected by a straight line to visualize the extrapolation
   predicted by the respective models. The vertical lines correspond to error
   bars in the solar system abundances taken from Goriely (1999).}  \label{F1}
\end{figure*}

Among the different thermodynamic parameters entering the canonical model, the
most critical one affecting the Th synthesis is obviously the maximum number of
neutrons captured by the initial seed nuclei, $n_{cap}^{max}$, which defines the
strong component of the r-process. In analogy with the s-process
nucleosynthesis, we can define a main r-process component responsible for the
production of all the elements up to the $A=195$ peak and part of the Pb
peak. This main component requires a value of $n_{cap}$ up to about 140. Till
now, there is no constraint from realistic models on the largest value that
$n_{cap}$ can take, i.e in analogy with the s-process, on a strong r-component
responsible for the bulk production of Pb and Bi. Considering canonical events
with values of $n_{cap}>140$ would lead to the production of the Pb-peak
elements, as well as Th, without affecting the synthesis of the lower-mass
elements. To illustrate such a sensitivity, multi-event calculations are
performed considering canonical events with a maximum number of neutrons
captured of $n_{cap}^{max}=$140, 145, 150 and 200 (Fig.~\ref{F1}). An excellent
fit is obtained for all isotopes with $A \lsimeq 204$ and seen not to be
affected at all by the change in the maximum value of $n_{cap}^{max}$
considered. On the contrary, increasing the $n_{cap}^{max}$ above 140 leads to
an increase in the production of the Pb-peak, Th and U elements. The large
uncertainties in the Pb-peak r-abundances cannot favour one or another fit.  The
complete absence of a strong r-component, i.e a maximum value of $n_{cap}=140$,
leads to a negative age (when derived from Eq.~\ref{eq1}) of the CS 22892-052
star, and can obviously be rejected. Including a strong r-component with
$n_{cap}=145$, 150 and 200 leads to a Th abundance that implies a star age
$T^{CS}_*=12.2$, 22.9 and 28.9 Gyr, respectively. Since the fit is constrained
by the upper value of the Pb and Bi abundances, increasing $n_{cap}^{max}$ above
200 does not affect the upper value of the Th abundance. More precisely, no
event with a value of $n_{cap} > 170$ contributes to the fit to the solar system
abundances with our adopted nuclear inputs.  In summary, any age below about 29
Gyr can thus be obtained just by adjusting the strength of such a strong
r-process component, unless the s- or r-origin of the Pb and Bi can be
determined with a greater accuracy.

\begin{figure*}
% POSTSCRIPT FILE: F2.ps
  
\resizebox{\hsize}{!}{\includegraphics[scale=.7]{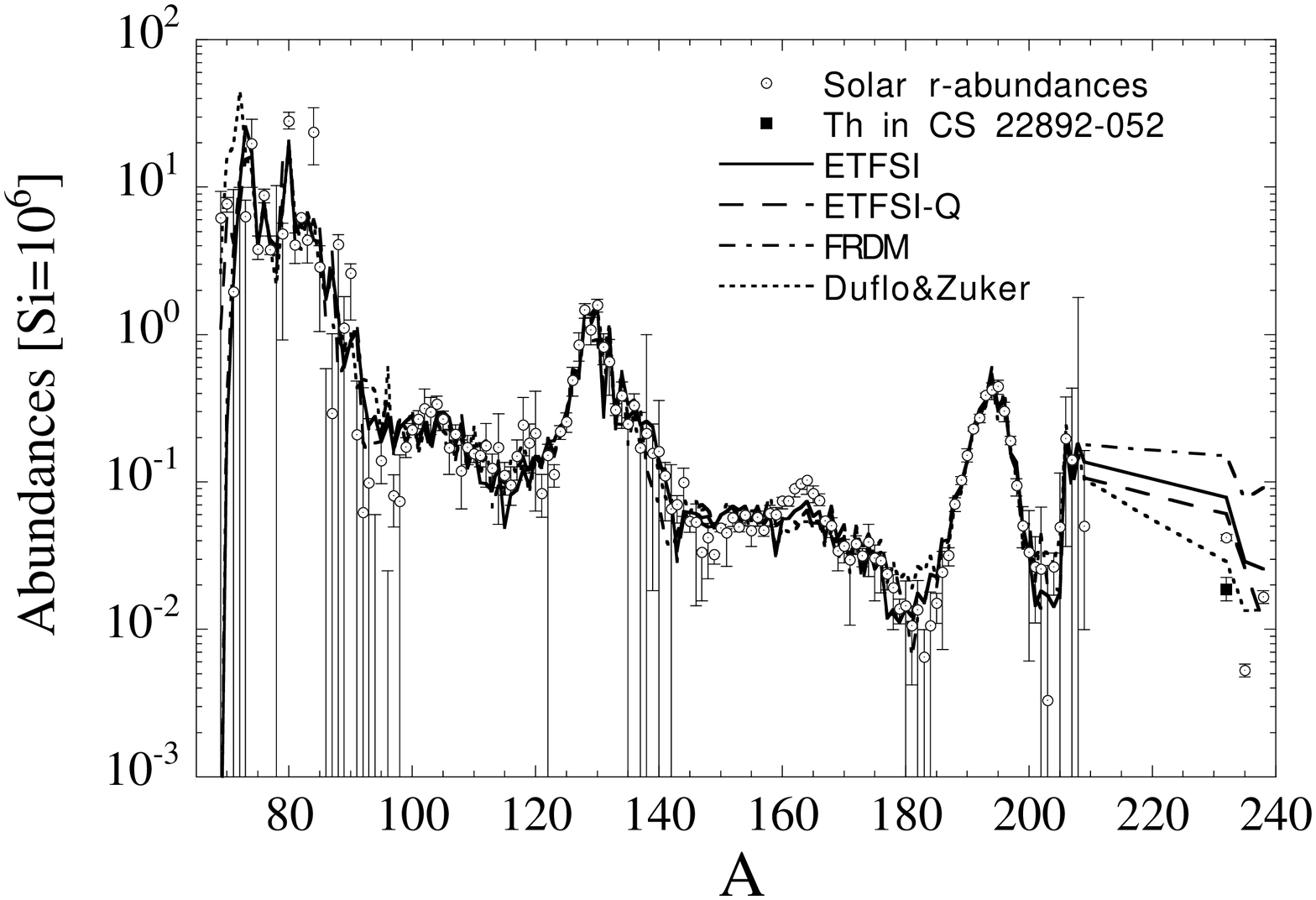}}
   \caption{ Same as Fig.~\ref{F1} where the predictions are obtained with the
   ETFSI (full), ETFSI-Q (dashed), FRDM (dot-dash) and Duflo \& Zuker (dotted)
   mass models. }  \label{F2}
\end{figure*}

\subsection{Sensitivity to nuclear physics input}
The most fundamental nuclear input to r-process models is well known to be the
nuclear masses. Various mass models are available, but for practical reasons we
only consider here in addition to the ETFSI model, the ETFSI-Q model (Pearson et
al. 1996) which takes into account the strong shell-quenching found in some
microscopic calculations on highly neutron-rich nuclei, the popular FRDM model
of M\"oller et al. (1995) and the recently-developed model of Duflo
\& Zuker (1995), hereafter DZ, based on a very different approach than the 
previously cited
models and which has proven its remarkable ability to predict experimentally
known masses. Many studies have compared the quality of these models and their
differences in the prediction of masses far away from the valley of
$\beta$-stability. Their impact on the r-process nucleosynthesis has also been
analyzed in various papers (e.g Goriely \& Arnould 1992), so that we will
restrict ourselves to analyze their respective predictions of the Th abundance.

Multi-event calculations are now performed making use of the 4 above-cited mass
models. The resulting fits are shown in Fig.~\ref{F2}. The fits to the stable
nuclei are of the same quality, and in particular in the Pb region, no major
differences in the predicted r-abundances can be observed. In particular, it
should be emphasized that no major deficiency in the fit is obtained in the
pre-peak regions at $A\simeq 120$ and $A\simeq 180$ whatever mass model is
used. Given the absence of realistic r-process models, there is obviously no
reason to favour one or another mass formula on grounds of parametric fits to
the solar r-abundance distribution, especially when dealing with the Th
abundance predictions which exclusively depend on the r-process paths in the $A
\ge 232$ region. The extrapolation to the Th abundance appears to be highly
affected by the mass model used. The estimate of the star age amounts to
$T_*^{CS}=28.9$, 23.8, 42.2 and 8.7 Gyr for the ETFSI, ETFSI-Q, FRDM and DZ
models, respectively. Such differences are not surprising, since it is well
known that the r-process paths for these mass models are significantly
different, in particular in the $Z>82$ region where the strength of the shell
correction energy around the $N=184$ shell closure can be very different. This
is not the case for the ETFSI and ETFSI-Q models, because the shell quenching
introduced in the ETFSI-Q model in the vicinity of the $N=184$ shell closure is
small.  Therefore, the abundance predictions in the Pb and actinide regions, and
consequently the stellar age predictions, are globally similar when making use
of ETFSI or ETFSI-Q. Compared with the other models, the DZ formula is
characterized by a steep slope of the mass parabola and a weak shell effect
around $N=184$, so that the progenitors responsible for the final Pb abundance
are found in a lower mass region by-passing partially Th and U. The Th abundance
obtained with FRDM model is higher than with the ETFSI mass models, because of a
more widely spread shell effect in the vicinity of the $N\le 184$ shell closure
affecting the r-process path down to $N=170$. The abundance peak around $N=184$
before freeze-out is consequently flattened to lower masses than in the ETFSI
case and is less affected by fission processes after freeze-out. A higher Th
abundance predicted with the FRDM model leads to a higher age estimate. The high
sensitivity of the predicted Th abundance to the mass model will not be resolved
before improving our mass predictions in the heavy ($Z>82$) neutron-rich region.

\begin{figure*}
% POSTSCRIPT FILE: F3.ps
  
\resizebox{\hsize}{!}{\includegraphics[scale=.65]{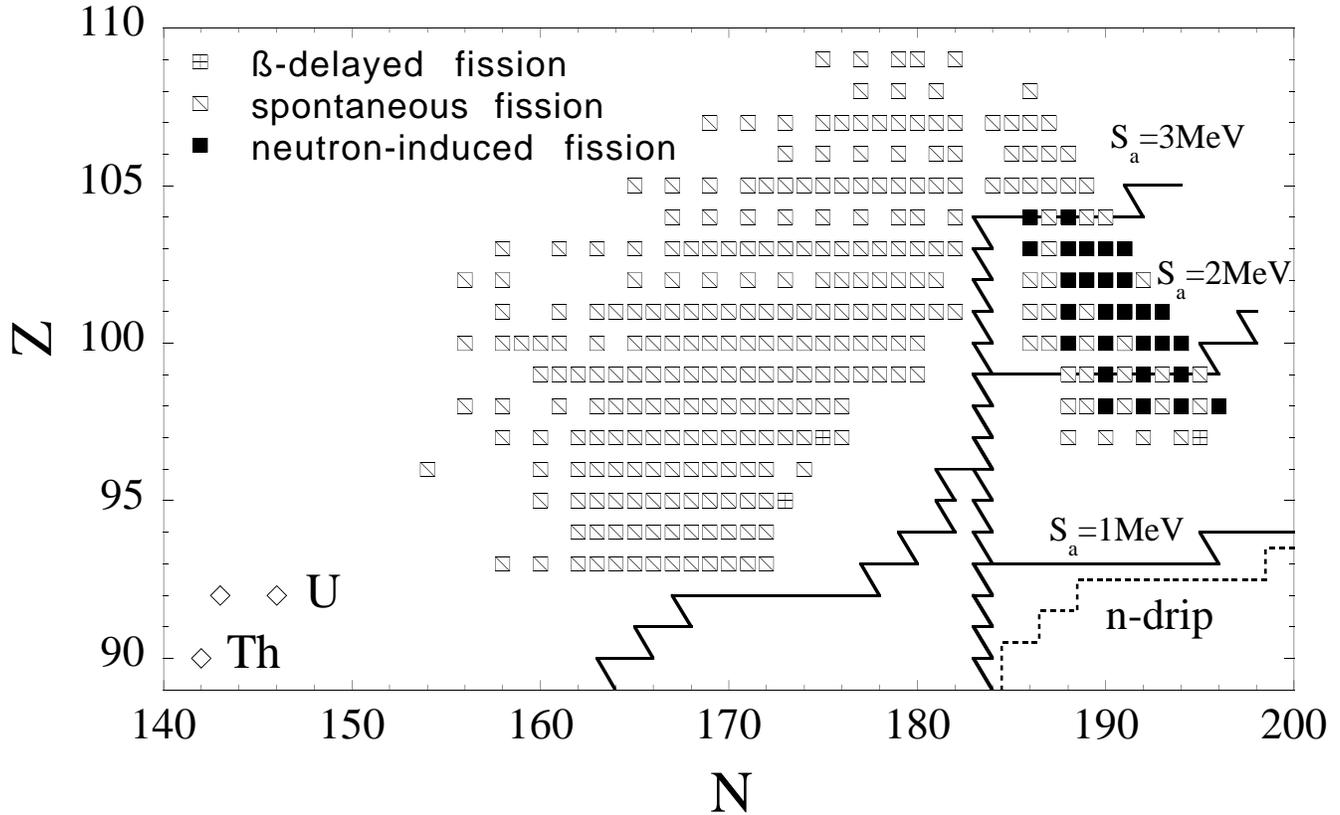}}
   \caption{ Representation in the nuclear chart of the dominating fission modes
   affecting the r-process flow, as given by the legend in the figure. Three
   r-process paths at $S_a=1$, 2 and 3 MeV (full line) are drawn for
   illustrative purposes, as well as the neutron drip line (dashed line).}
   \label{F3}
\end{figure*}

Another important ingredient in the Th nucleosynthesis concerns the fission
processes, i.e the spontaneous, neutron-induced and $\beta$-delayed
fission. Most of the r-pro\-cess calculations do not include the fission
processes at all or only partially. However, when dealing with Th
cosmochronometry, fission processes must be included in the most careful way in
order to describe the competing processes responsible for the final Th abundance
(namely $\alpha$-decays, $\beta$-decays and fissions) correctly. The recent
large scale calculation of ETFSI fission barriers (Mamdouh et al. 1998) up to
$A=295$ is used to test the significance of fission on the Th
cosmochronometry. The spontaneous, neutron-induced and $\beta$-delayed fission
probabilities are determined in the same way as in Kodoma \& Takahashi
(1975). It should, however, be stressed that when not available experimentally,
the spontaneous fission rates are derived from a new regression fit to
experimental data based on our fission barrier predictions. Figure~\ref{F3}
shows the nuclear regions where the different fission modes influence the
r-process flows. Because of the strong ETFSI shell effect on the fission
barriers around $N=184$, no fission recycling is found during the neutron
irradiation, at least before crossing the $N=184$ closure. Only r-process paths
characterized by an astrophysical parameter $S_a\gsimeq 2~{\rm MeV}$ (for more
details about the astrophysical parameter, see Goriely \& Arnould 1992) are
stopped by neutron-induced fission. Spontaneous fission can also affect such
r-process paths before neutron freeze-out. $\beta$-delayed fission is found to
be of small importance compared with the other decaying modes, even after
freeze-out. On the contrary, the spontaneous fission is found to be faster than
the $\beta$-decay for almost all isobaric chains above $A=250$ and crucial in
estimating the final r-abundances in the Pb and actinide region. In the specific
fits studied in the present paper, the fission fragments do not affect the
low-mass abundance distribution. Obviously, all the above conclusions should be
taken with care, because of the uncertainties remaining in the determination of
the fission barriers and fission probabilities, which are to be studied in a
forthcoming paper.

\begin{figure*}
% POSTSCRIPT FILE: F4.ps
  
\resizebox{\hsize}{!}{\includegraphics[scale=.7]{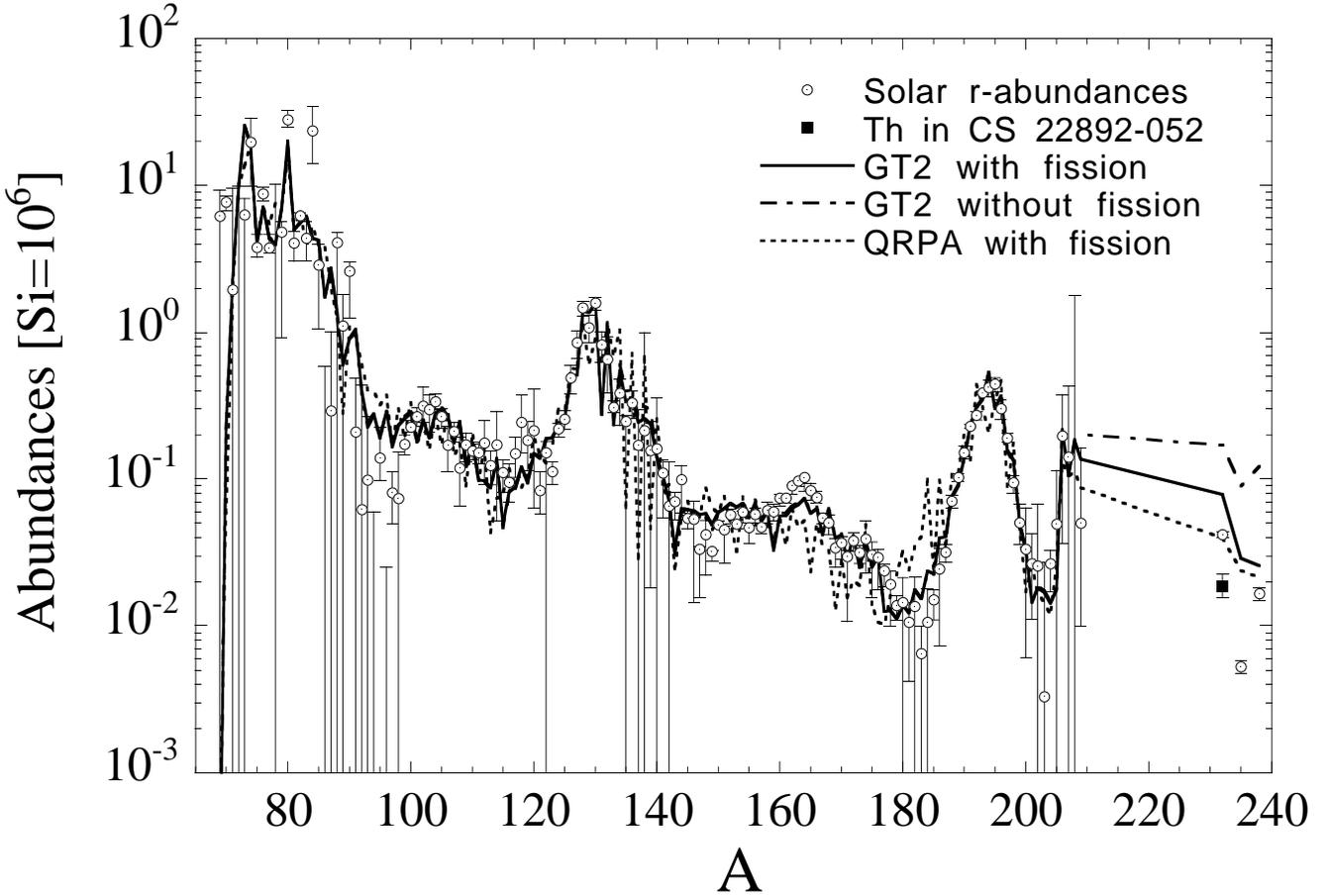}}
   \caption{ Same as Fig.~\ref{F1} where the predictions are obtained using the
   GT2 $\beta$-decay and $\beta$-delayed neutron emission rates with (full) or
   without (dash-dot) fission processes. The dotted curve corresponds to the use
   of the QRPA $\beta$-decay and $\beta$-delayed neutron emission rates
   including fission.}  \label{F4}
\end{figure*}

In order to quantify the impact of the fission processes on the Th
cosmochronometry, a multi-event calculation is reiterated switching off all the
fission processes. This numerical test just aims at illustrating the largest
error possibly made when neglecting fission, but should not be regarded as a
sensible test case for the Th prediction. It is found that the neglect of
fission gives rise to an increase of the age of CS 22892-052 by 12.3 Gyr on
grounds of the abundance distribution shown in Fig.~\ref{F4}.  It can also be
seen that when including fission processes, the fission fragments do not modify
the global abundance distribution.  Obviously, a complete and consistent
treatment of the fission processes (especially spontaneous fission) is required
to build a reliable cosmochronometry on the actinides.

Figure~\ref{F4} also presents uncertainties associated with $\beta$-decays and
$\beta$-delayed neutron emission by comparing the solar fits obtained with the
GT2 model and the QRPA model of M\"oller et al. (1997).  Both models have been
extensively used in previous works dedicated to the r-process nucleosynthesis,
so that it is of interest to study their influence on the Th
cosmochronometry. If use is made of the QRPA model instead of the GT2 model, a
reduction from 28.9 Gyr down to 15.1 Gyr is obtained for the age of CS
22892-052.Once again, it should be added that although the fit to the solar
distribution obtained with the QRPA model is slightly worse than the one
obtained with the GT2 approach, it cannot be rejected {\it a priori}, since
other nuclear or astrophysics shortcomings of the model can be responsible for
the observed discrepancies (for example in the $A=180$ region). As stated
previously, given our poor understanding of the r-process nucleosynthesis
(especially of the astrophysical site) the quality of nuclear models should not
be tested on astrophysics arguments like fits to the solar abundance
distribution.

\subsection{Uncertainties in the solar r-abundance}
\begin{figure*}
% POSTSCRIPT FILE: F5.ps
  
\resizebox{\hsize}{!}{\includegraphics[scale=.7]{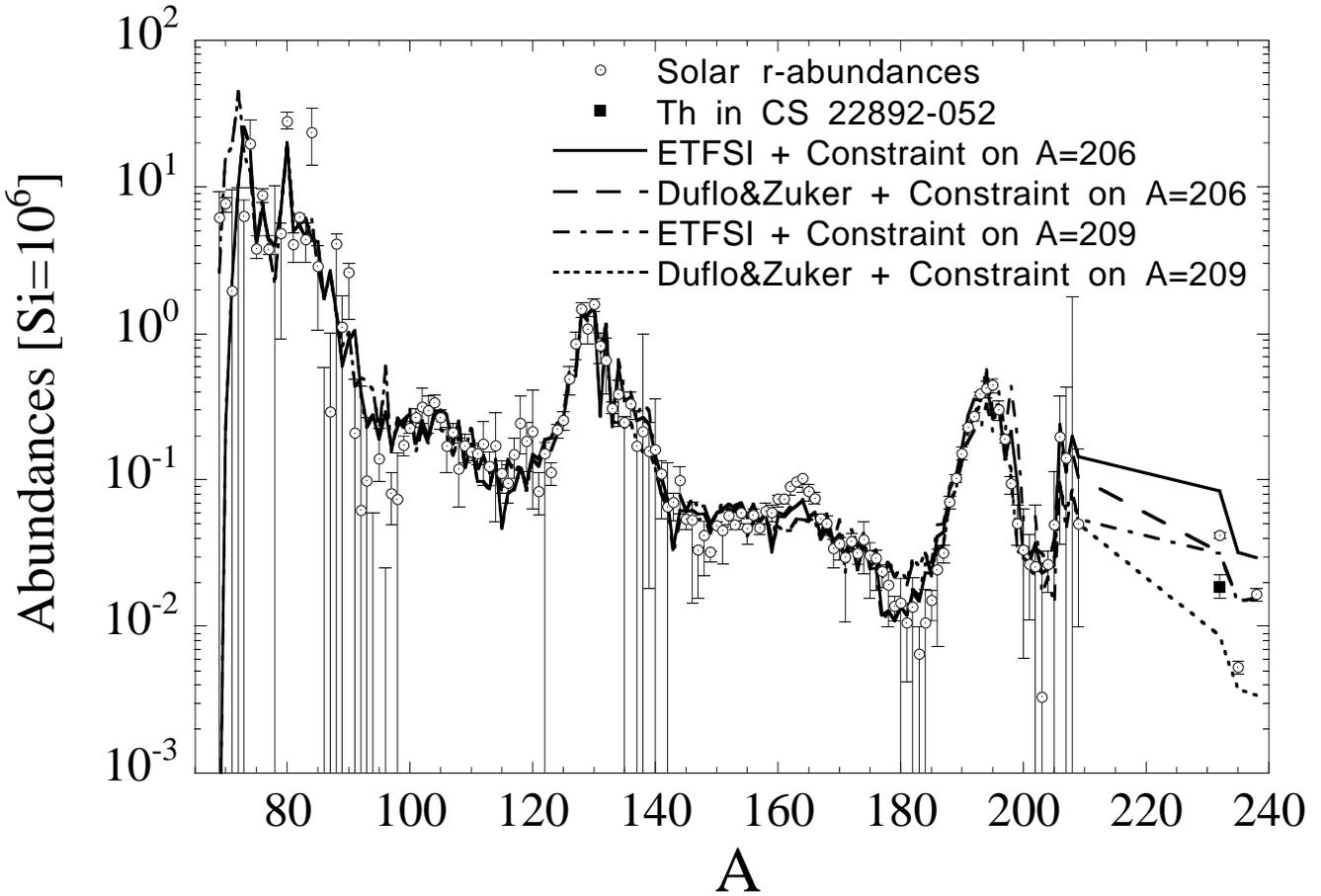}}
   \caption{ Same as Fig.~\ref{F1} when the fit is constrained on $^{204}{\rm
   Hg}$ and $^{206}{\rm Pb}$ solar abundances with the ETFSI (full) or DZ
   (dashed) mass models. The dash-dot and dotted lines correspond to the
   constrain on $^{204}{\rm Hg}$ and $^{209}{\rm Bi}$ abundances with ETFSI and
   DZ mass models, respectively. } \label{F5}
\end{figure*}

The uncertainties still affecting the s-process model are responsible for
non-negligible imprecisions in the determination of the residual
r-abun\-dan\-ces of the solar system content (Goriely 1999), in particular for
the so-called s-dominant isotopes and Pb-Bi isotopes (see the large error bars
in Fig.~\ref{F1}). The principal source of uncertainty in the solar r-abundance
of Pb and Bi lies in our ignorance of the relative s- and r-contribution to
their production.  Both processes can produce Pb-Bi almost entirely, so that the
r-contribution cannot be estimated in a reliable way on grounds of s-process
calculations. This problematic aspect of the solar abundance splitting in the Pb
region can only be resolved through realistic modelling of the s-process (or
accurate abundance determination of Pb at the surface of ultra-metal-poor stars
provided the assumption of the r-process universality be confirmed). It should
be kept in mind that the prediction of the Th and Pb-Bi abundances are strongly
correlated, so that any uncertainty in the solar r-abundances of the Pb and Bi
elements is translated into the exponentially dependent uncertainty in the star
age. Among the $A>200$ r-isotopes, $^{204}{\rm Hg}$, $^{206}{\rm Pb}$ and
$^{209}{\rm Bi}$ play an important role since their r-abundances are better
determined than for their neighbours.  The solar r-abundance of $^{204}{\rm Hg}$
is indeed well determined, so that it seems logical to constrain the fit in such
a way as to reproduce the $^{204}{\rm Hg}$ solar r-abundance. As regards
$^{206}{\rm Pb}$ and $^{209}{\rm Bi}$, they are characterized by a relatively
well-determined abundance to which the Th production is directly correlated.
However, reproducing the recommended solar abundance of $^{204}{\rm Hg}$,
$^{206}{\rm Pb}$ and $^{209}{\rm Bi}$ simultaneously appears to be impossible
without strongly deteriorating the fit to the $A=195$ peak.  For this reason, we
reiterate multi-event calculations in which the fitting procedure is constrained
(with an extra statistical weight) to the solar abundance of $A=204$ and 206 in
one case and $A=204$ and 209 in the other case, using two different mass models,
namely the ETFSI and DZ models (Fig.~\ref{F5}). Although a negative age is
obtained with the DZ masses when constraining the fit to $^{209}{\rm Bi}$, the
other predicted ages are 30.4 Gyr for the ETFSI calculation constrained to
$^{206}{\rm Pb}$ and 10.5 Gyr in the two remaining cases.  So, in addition to
the large sensitivity of the stellar age to the mass models as studied in the
previous section, the uncertainties in the solar r-abundances appear to affect
the age determination by about 20 Gyr.  As long as the s- or r-origin of the Pb
and Bi solar abundance is not determined with high accuracy, the Th
cosmochronometry will not provide any reliable age estimate.

\section{The age of the stars}
\begin{figure*}
% POSTSCRIPT FILE: F6.ps
   \resizebox{\hsize}{!}{\includegraphics[angle=-90]{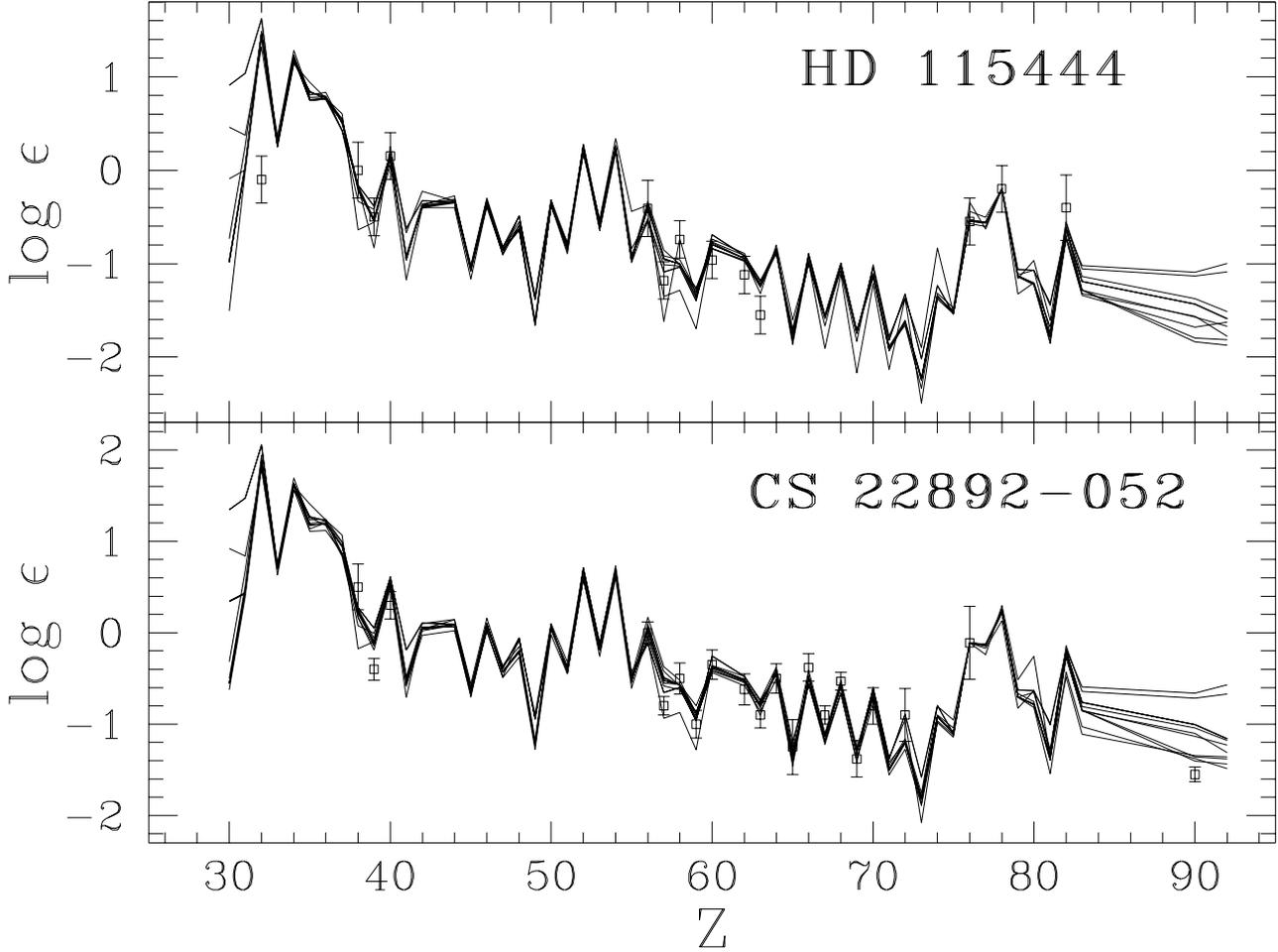}}
   \caption{Elemental abundance distributions observed in HD 115444 and CS
   22892-052 (open squares) compared with multi-event r-abundance predictions
   (see text).}  \label{F6} \end{figure*}

So far, we have been evaluating the stellar ages on the basis of a fit to the
solar r-abundance distribution. Such a procedure is obvioulsy based on our
initial fundamental assumption that the r-process site is unique in the
Galaxy. This basic assumption is a fundamental prerequisite to build a Th
cosmochronometry upon the abundance analysis of metal-poor stars at the present
time, since, as explained above, a direct fit to the abundance distribution
observed at the surface of metal-poor stars would present even larger
uncertainties, because of the restricted number of elements observed, the
impossibility to distinguish isotopic ratios and the much smaller accuracy in
the abundance data as compared with our solar system. However, in the specific
case of Pb, the uncertainties still affecting the solar abundance might be
greater than the ones found in the observation at the surface of HD 115444 or HD
126238 (Sneden et al.  1998), so that the Pb abundance in such stars could
probably be more constraining on the Th predictions than the solar
value. Unfortunately, the Th abundance has not been determined yet in such
stars, so that in this case, we limit the discussion on the relative stellar
age.  Compared with CS 22892-052, only a small number of r-elements are observed
in HD 115444, but it is at the moment the only metal-poor star in which Pb and
Th lines are detected simultaneously.  All the physically sound (i.e leading to
a positive age $T^{CS}_*$) calculations presented in the previous section are
compared, in Fig.~\ref{F6}, with the elemental abundance distribution observed
in HD 115444 and CS 22892-052. We only retain calculations which provide a good
fit to the observed $Z \ge 55$ abundances, and in particular to the Pb abundance
in the case of HD 115444.  In order to achieve a good fit to all elements
observed and to optimize the extrapolation to the Th region, a normalization of
the abundance curves is done on the heaviest elements accurately observed, i.e
Pt and Os for HD 115444 and CS 22892-052, respectively. In this case, the age of
CS 22892-052 is found to lie in the $7 \le T^{CS}_* {\rm[Gyr]} \le 39$ range,
and a similar error range of about 32 Gyr is predicted for the age of HD
115444. The different predictions of the stellar age, as well as the elemental
abundances of Eu, Pb, Bi, Th and U are summarized in Table~\ref{tab1}.  The
stellar age can be determined in two different ways. If normalized to the
calculated Eu abundance, the age $T_*^1$ has the advantage of beeing free from
uncertainties in the normalization procedures on the observed abundances, but is
sensitive to the theoretical uncertainties made in the r-process predictions. In
particular, it is well known that the origin of the r-nuclides in the
$A\simeq160$ region is not easily explainable (e.g Meyer \& Mullenax 1998). The
impact of such errors remains to be estimated. On the contrary, normalizing the
calculated Th abundance on the observed Eu avoids theoretical complications
inherent in the origin of the $A\simeq160$ nuclides, but gives rise to
additional errors of the order of $\pm 0.1$ dex on the abundances (as seen in
Fig.~\ref{F6}), i.e of $\pm 4.7$ Gyr on the age $T_*^2$.
%-----------------------------------------------------------------------------
%-----------------------------------------------------------------------------

%-------------------------------------------------------------------------
% Abundances for HD 115444 and CS 22892-052 
%  TABLE
%\small
\begin{table*}
  \centering \caption{Ages $T_*$ (in Gyr) of CS 22892-052, elemental abundances
  (in log$\epsilon$) for Eu, Pb, Bi, Th, U and log(Th/U) predicted by the
  different theoretical calculations shown in Fig.~\ref{F6} and analyzed in
  Sect.~3. $T_*^1$ ($T_*^2$) corresponds to the age calculated with a Th
  abundance normalized to the calculated (observed) Eu abundance. }

   \begin{tabular}{|l|cccccccc|} \hline Comment (see Sect.~3)& Eu & Pb & Bi & Th
   & U & Th/U & $T_*^1$ & $T_*^2$ \\ \hline Standard ($n_{cap}^{max} = 200$,
   ETFSI, GT2) & -0.79 & -0.20 & -0.76 & -1.00 & -1.16 & 0.16 & 22.9 & 25.5 \\
   $n_{cap}^{max} = 145$ & -0.82 & -0.43 & -1.03 & -1.38 & -1.49 & 0.11 &6.6 &
   8.2 \\ $n_{cap}^{max} = 150$ & -0.80 & -0.27 & -0.85 & -1.13 & -1.23 & 0.10
   &17.3 & 19.5 \\ ETFSI-Q masses & -0.79 & -0.26 & -0.86 & -1.10 & -1.32 & 0.22
   &18.2 & 21.0 \\ FRDM masses & -0.91 & -0.14 & -0.64 & -0.72 & -0.67 & -0.05 &
   41.6 & 38.9 \\ DZ masses & -0.75 & -0.19 & -0.85 & -1.40 & -1.44 & 0.04 & 2.4
   & 6.9 \\ ETFSI + constraint on $A=206$ & -0.86 & -0.22 & -0.81 & -1.04 &
   -1.18 & 0.14 & 24.3 & 23.8 \\ DZ + constraint on $A=206$ & -0.75 & -0.20 &
   -0.85 & -1.36 & -1.38 & 0.02 & 4.3 & 8.9 \\ ETFSI + constraint on$A=209$ &
   -0.74 & -0.48 & -1.11 & -1.34 & -1.36 & 0.02 & 4.7 & 9.6 \\

  \hline
\end{tabular}
\label{tab1}
\end{table*}
%\normalsize
%-------------------------------------------------------------------------

Although the present study suggests that at the moment great care should be
taken in estimating the age of the stars on the basis of observed Th abundance,
new accurate observations of heavy r-elements could put the Th cosmochronometry
on safer grounds, especially if Th and U lines could be observed accurately and
simultaneously in metal-poor stars, as already stressed by Arnould \& Takahashi
(1999). As a matter of fact, if the Th and U lines are available, an age
estimate could be derived from the expression
\begin{equation}
\log \Bigl({{\rm Th} \over {\rm U}} \Bigr)_{obs} = \log \Bigl({{\rm Th} 
\over {\rm U}}
\Bigr)_{r} +\log {\rm e} ~ \Bigl( {1 \over \tau({\rm U})} - {1\over 
\tau({\rm Th})}\Bigr)~T_*
\label{eq2}
\end{equation}

\noindent where $\tau({\rm U})=6.41~{\rm Gyr}$ is the characteristic 
$\alpha$-decay
time\-scale of U. As seen in Fig.~\ref{F6} and Table 1, $\log({\rm Th/U})_r$ is
found to lie within an 0.1 range, whatever theoretical inputs are used in the
r-process model. Such an accurate estimate is principally bound to the fact that
Th and U are neighbour nuclei, and consequently their production ratio is not
strongly affected by unreliable extrapolation procedures, but rather by local
nuclear uncertainties, such as nuclear masses or fission processes in the
actinide region.

\begin{figure}
% POSTSCRIPT FILE: F7.ps
   \resizebox{\hsize}{!}{\includegraphics[scale=.32]{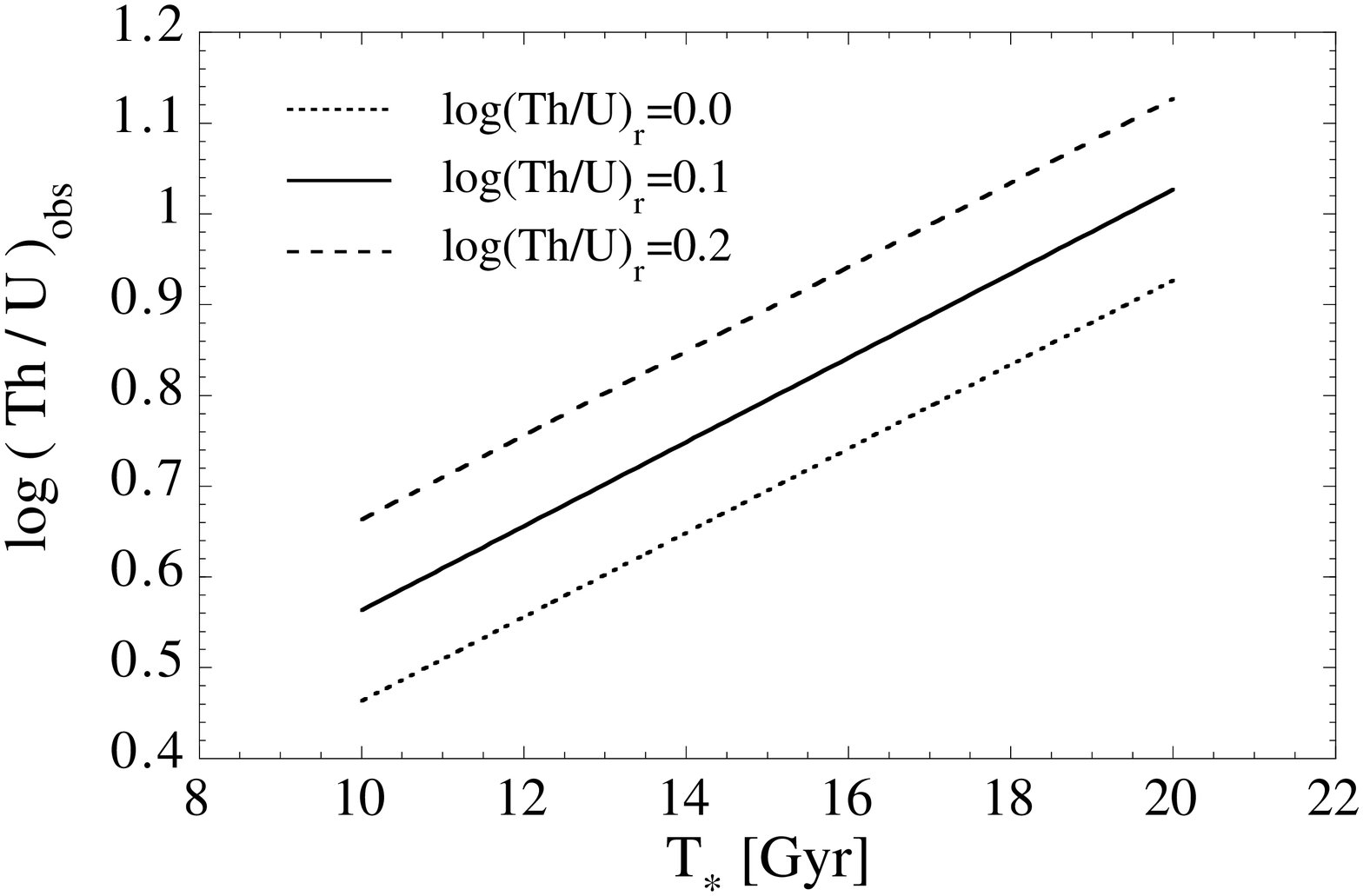}} \caption{
   Relation between the observed Th/U ratio and the age of the star $T_*$ for 3
   different estimates of the production ratio.}
\label{F7}
\end{figure}

 From Eq.~(\ref{eq2}), it is found that a $\pm 0.1$ error on the observed or
 predicted production ratio of Th/U gives rise to a $\pm 2.1$ Gyr error on
 $T_*$. A future simultaneous observation of Th and U lines in ultra-metal-poor
 stars could therefore open the way to an accurate age determination in contrast
 to the still complicated Th cosmochronometry based on Th lines
 only. Equation~(\ref{eq2}) is shown in a graphical form in Fig.~\ref{F7} for 3
 possible values of the Th/U production ratio. Note, however, that a deeper
 analysis of the Th/U production ratio would be required before rushing into an
 age determination. In particular, uncertainties in the fission processes have
 not been included in the present study, but could possibly affect the predicted
 Th/U ratio.

%------------------------------------------------------------------------------
%-------------------------------------------------------------------------------

\section{Conclusions}

The present paper analyzes critically the impact of the remaining uncertainties
in the existing r-process models on the determination of the age of the Galaxy
derived on grounds of the Th cosmochrometry. Although the direct and accurate
observation of the Th abundance in ultra-metal-poor stars provides a possible
method to date the age of the star without calling for complex model of the
chemical evolution of the Galaxy, the Th cosmochronometry remains affected by
all the difficulties associated with our poor understanding of the r-process
nucleosynthesis. Even if we disregard the difficulty associated with the
fundamental assumption relative to the unicity of the r-process site in the
Galaxy, it is impossible at the present stage reliably to determine the age of a
ultra-metal-poor star from parametric models of the r-process. The major
difficulties lie in the very s- or r-origin of the Pb and Bi isotopes, in
addition to the unavoidable problem related to the prediction of nuclear data
(principally masses) and the still unknown thermodynamic conditions in which the
r-process takes place (in particular concerning the maximum neutron irradiations
that can be reached). New accurate observations including Pb and Th lines in
ultra-metal poor stars could shed light on the ability of the r-process to
produce the Pb-peak elements, and consequently constrain the Th
cosmochronometry. In particular, the Th cosmochronometry could be put on safer
grounds, especially if Th and U lines could be observed accurately and
simultaneously in ultra-metal-poor stars. As regards nuclear uncertainties, only
theoretical as well as experimental effort to come can help us achieve the
required accuracy to pretend to derive the age of the Galaxy on the basis of the
actinides cosmochronometry.

\begin{acknowledgements}                                                                                                          
The authors are grateful to M. Arnould for helpful discussions
\end{acknowledgements}


\begin{thebibliography}{}

\bibitem[ ]{ } Aboussir Y., Pearson J.M., Dutta A.K., Tondeur F., 1995, At. Data
Nucl. Data Tables 61, 127
\bibitem[ ]{ } Arnould M., Takahashi K., 1999, Rep. Prog. Phys. 62, 393
\bibitem[ ]{ } Bouquelle V., Cerf N., Arnould M., Tachibana T., Goriely S., 1996,
A\&A 305, 1005
\bibitem[ ]{ } Cowan J.J., Mc William A., Sneden C., Burris D.L., 1997, ApJ 480, 
246
\bibitem[ ]{ } Duflo J., Zuker A.P., 1995, Phys. Rev C52, 23
\bibitem[ ]{ } Goriely S., Arnould M., 1996, A\&A  312, 327
\bibitem[ ]{ } Goriely S., Arnould M., 1997, A\&A  322, L29
\bibitem[ ]{ } Goriely S., 1999, A\&A  342, 881
\bibitem[ ]{ }  Kodoma T., Takahashi K., 1975, Nucl. Phys. A239, 489
\bibitem[ ]{ }  Mamdouh A., Pearson J.M., Rayet M., Tondeur F., 1998, Nucl. Phys.
A644, 389
\bibitem[ ]{ } Meyer B.S., Mullenax D.J. 1998, Nuclei in the Cosmos dol V, 
eds N. Prantzos, S.  Harrisopoulos (Gif-sur-Yvette: Editions Fronti\`eres) p. 289
\bibitem[ ]{ } M\"{o}ller P., Nix J.R., Myers  W.D., Swiatecki W.J., 1995, At.
Data Nucl. Data Tables 59, 131
\bibitem[ ]{ } M\"{o}ller P., Nix J.R., Kratz K.-L., 1997, At. Data
Nucl. Data Tables 66, 185
\bibitem[ ]{ } Pearson J.M., Nayak R.C., Goriely S., 1996, Phys. Lett. B 387, 455
\bibitem[ ]{ } Pfeiffer B., Kratz K.-L., Thielemann F.-K. et al., 1998, In Proc. 
of the 9th Workshop on Nuclear Astrophysics, eds W. Hillebrandt, E. M\"uller (MPA,
Munich) p.168
\bibitem[ ]{ } Seeger P.A., Fowler W.A., Clayton D.D., 1965, ApJS 11, 121
\bibitem[ ]{ } Sneden C., Mc William A., Preston G.W., et al., 1996, ApJ 467, 819
\bibitem[ ]{ } Sneden C., Cowan J.J., Burris D.L., Truran J.W., 1998, ApJ 496, 235
\bibitem[ ]{ } Tachibana T., Yamada M., Yoshida Y., 1990, Progr. Theor.
Phys. 84, 641

\end{thebibliography}
\end{document}